\documentclass[superscriptaddress,amsmath,10pt,floatfix,prd,showpacs,preprintnumbers]{revtex4}
\usepackage[mathscr]{eucal}
\usepackage[dvips]{color}
\usepackage[dvips]{graphicx}
\usepackage{epsf}
\usepackage{bm}
\usepackage{amsmath}
\usepackage[normalem]{ulem}
\usepackage{graphicx}

\def\){\right)}
\def\({\left(}
\def\]{\right]}
\def\[{\left[}

\newcommand{\be}{\begin{equation}}

\newcommand{\ee}{\end{equation}}
\newcommand{\bea}{\begin{eqnarray}}
\newcommand{\eea}{\end{eqnarray}}

\newcommand{\threeier}[1]{\int {d\!\!\!^{-3}#1}}
\newcommand{\intspace}[1]{\int d^4 {#1}\,}
\newcommand{\ha}{\frac{1}{2}}
\newcommand{\bfr}{{\bf{r}}}
\newcommand{\bfp}{{\bf{p}}}
\newcommand{\bfk}{{\bf{k}}}

\newcommand{\tepsp}{\tilde{\epsilon}({\bf{p}})}
\newcommand{\xip}{\xi({\bf{p}})}
\newcommand{\txip}{\tilde{\xi}({\bf{p}})}
\newcommand{\dmu}{\delta\mu}
\newcommand{\tdmu}{\delta\tilde{\mu}}
\newcommand{\dm}{\delta\mu}
\newcommand{\tdm}{\delta\tilde{\mu}}
\newcommand{\tmu}{\tilde{\mu}}
\newcommand{\dn}{\delta n}
\newcommand{\tr}{{\rm{Tr}}}

\newcommand{\Pd}{\Psi^\dagger}
\newcommand{\hatp}{\hat{\bf{p}}}

\begin{document}

\title{A Mean Field Analysis of Pairing in Asymmetric Fermi Systems at Finite Temperature}
\date{\today}
\author{Rishi~Sharma}
\email{rishi@lanl.gov}
\affiliation{Theoretical Division, Los Alamos National Laboratory,\\
 Los Alamos, NM 87545, USA.}
\author{Sanjay~Reddy}
\email{reddy@lanl.gov}
\affiliation{Theoretical Division, Los Alamos National Laboratory,\\
 Los Alamos, NM 87545, USA.}

\preprint{LA-UR-08-2124}

\begin{abstract}
We study the phase diagram of a two component Fermi system with a weak attractive
interaction.  Our analysis includes the leading order Hartree energy shifts and
pairing correlations at finite  temperature and chemical potential difference
between the two fermion species.   We show that in an asymmetric system, the
Hartree shift to the single particle energies are important for the phase
competition between normal  and superfluid phase and can change the phase
transition curve qualitatively.  At large asymmetry we find that a novel but
somewhat fragile superfluid state can be favored due to finite temperature
effects.   We also investigate the transition between the normal phase and an
inhomogeneous superfluid phase to study how gradient instabilities evolve with
temperature and asymmetry. Finally, we adopt our analysis to study the density
profiles of similar asymmetric Fermi systems that are being observed in cold
atom experiments.    
\end{abstract}
\pacs{74.25.Dw,74.20.Fg} 
\maketitle

\section{Introduction}
The superfluid nature of the ground state of symmetric two-component Fermi
systems, where pairing occurs between equal densities of the two species, has
been well established, both on theoretical and experimental grounds, since the
pioneering work of Bardeen, Cooper and Schrieffer (BCS) fifty years ago
\cite{BCS:1957}. However, the phase structure of asymmetric Fermi systems still
remains unclear. In particular, theoretical work suggests several novel and
competing superfluid phases may be possible. These include: (i) the Ferrel,
Fulde, Larkin, Ovchinnikov (FFLO) phase, which is an inhomogeneous superfluid
phase with a spatially varying superfluid order parameter \cite{FF:1964,LO:1965};
(ii) the breached-pair or gapless superfluid phase, which is a homogeneous
superfluid phase containing gapless fermionic excitations even at zero
temperature \cite{Forbes:2005}; and (iii) phases with broken rotational symmetry
where pairing is facilitated by a deformation of the Fermi surface
\cite{Muther:2002,Sedrakian:2005}.  Quantum Monte Carlo (QMC) calculations 
have been used to calculate the thermodynamic properties of a symmetric system at
zero temperature \cite{Carlson:2005} for arbitrary interaction strength. The 
authors of \cite{Son:2006} used these results to deduce the presence of a
splitting point in the phase diagram of asymmetric systems in the strong
interaction regime, where a gapped superfluid phase, a gapless superfluid phase
and an inhomogeneous superfluid phase coexist. Their analysis suggests that with the
interaction strength tuned to a value such that the $s$-wave scattering length between the two
species is infinite, on increasing the difference in the chemical potentials of
the two species the system undergoes a first order transition from a homogeneous 
gapped superfluid phase to an inhomogeneous phase, before eventually transiting to the normal phase. 

To investigate the possible existence of these phases, cold atom experiments are
now exploring the thermodynamic and linear response properties of asymmetric
Fermi systems with attractive interactions
\cite{Partridge:2006a,Shin:2006,Partridge:2006b,Shin:RF2007,shin:2007}. These
experiments which trap and cool two hyperfine states of $^6$Li atoms, have
unprecedented control over the sample. They can: (i) magnetically tune the
interaction strength between the two hyperfine states through Feshbach
resonances; (ii) control the population asymmetry by loading different numbers
of atoms; and (iii) vary the temperature. In the strongly interacting regime,
where the two-body scattering length is large, these experiments have already
observed how the superfluid properties change with number asymmetry and
temperature. Experimental measurements of the density profiles and the response
to radio frequency probes seem to indicate that these novel phases are not
realized in cold atom traps. Instead one finds strong indications of a
first-order phase transition between a superfluid state with zero number
asymmetry and a normal state with a large asymmetry
\cite{Partridge:2006a,Shin:2006}. 

The absence of intervening novel superfluid phases in cold atom experiments is
intriguing. These phases may still exist at very low temperature and in the weak
coupling regime.  To address if these phases can be realized in experiments in
the future, we  investigate the phase structure of asymmetric Fermi systems at
finite temperature and establish the parameter region where interesting new
phases of superfluidity can be realized. Our study differs from similar
investigations reported in Ref.~~\cite{parish-2007,gubbels-2006-97} in two ways.
(For a discussion of the phase structure of asymmetric systems at finite
temperature in condensed matter systems,
see~\cite{sarma-1963,1994AnP...506..181B, buzdin-1997,combescot-2004-68}.) 
First, we restrict our analysis to the weak coupling region and perform a
self-consistent calculation of the thermodynamic properties of both the normal and
the superfluid phases within the purview of mean field theory (Hartree
approximation). One drawback of this approach is that we neglect particle-hole
screening which is important in the gap equation at weak
coupling~\cite{Gorkov:1961,Heiselberg:2000}. This screening is expected to
reduce the gap by a factor $\simeq 2.2$.  In the conclusions
(section~\ref{section:conclusions}) we will return to a discussion of how this
suppression may affect our final answers.  Second, we establish the region in
temperature and number asymmetry where the homogeneous states are unstable with
respect to small amplitude perturbations.    

 Although our analysis is strictly valid only in the weak coupling regime, unlike
earlier work, we properly account for the single particle energy shifts -- or
Hartree corrections -- that are present both in the normal and superfluid state,
in a self-consistent field theoretic approach. We find that these corrections
can change the location and shape of the phase boundaries on the phase diagram
of the two component Fermi gas. We also find that a new gapless, finite
temperature superfluid state is favored over the normal state at temperatures
above the critical temperature for the first-order transition. In a narrow
window of chemical potential asymmetry, with increasing temperature there is
first a first-order transition from the superfluid to the normal phase and then
at higher temperature a second-order transition to a weakly superfluid state
which exists in limited temperature interval. In this state, finite temperature
effects that smear the Fermi surface, facilitate pairing.  We will refer to this
homogeneous phase as the ``fragile'' superfluid and it has lower energy than the
normal state but is unstable with respect to gradient instabilities.

To study these instabilities we expand the free energy in a Ginzburg-Landau
series about the solution of the gap equation, and calculate the coefficient
of the quadratic term in the order parameter as a function of its Fourier mode
index. We ask the question whether this coefficient is negative for some Fourier
mode(s), indicating a gradient instability.  The simplest
case is an expansion about the normal phase with a zero background value for 
the difermion condensate. In the presence of a small
position dependent condensate 
\begin{equation}
\eta(\bfr) = \threeier{\bfk}e^{i\bfk\cdot\bfr}\eta(\bfk) \;,
\end{equation}
we write the change in the free energy as,
\begin{equation}
\threeier{\bfk} \bigl(\alpha +
f(|\bfk|)\bigr)\eta(\bfk)\eta(-\bfk)+{\cal{O}}(\eta^4)\;,
\end{equation}
where we have split the terms in a way that $f(|\bfk|)$ is zero for $\bfk=\bf0$.
$\alpha<0$ points to an instability towards the formation of a homogeneous
condensate while $f(|\bfk|)<0$ points to an instability towards an inhomogeneous
modulation of the condensate. We find that there is a window of temperatures 
and chemical potential splitting where the inhomogeneous superfluid phases 
may be favored.

The plan of the paper is as follows. We begin by writing down the lagrangian,
and reviewing how to include Hartree corrections to the calculation of the gap
parameter $\Delta$ and the free energy $\Omega$ of the system in
section~\ref{section:model lagrangian}. We then proceed with the calculation of
$\Delta$ and $\Omega$ in section~\ref{section:phase boundary} which will let us
find the phase boundary between the superfluid and the normal phase. We first
consider homogeneous phases, meaning $\Delta$ independent of position, and
follow with a discussion of inhomogeneous phases. We conclude with a summary of
our results in section~\ref{section:conclusions} where we use our expressions
to calculate the polarization as a function of the distance from the center of
an isotropic atomic trap, for typical trap parameters. We also identify the region
where inhomogeneous phases may be found in a typical trap geometry.

\section{The model lagrangian \label{section:model lagrangian}}

 We describe the gas of two species of fermions, $\psi_1$ and $\psi_2$, with
chemical potentials $\mu_1$ and $\mu_2$, at finite temperature $T$, by a model lagrangian 
density of the form,
\begin{equation}
{\cal{L}} = \psi_\alpha^\dagger\Bigl(
  (i\partial_t-\xi(\hatp))\delta_{\alpha\beta} + \delta\mu\sigma^3_{\alpha\beta}
  \Bigr)\psi_\beta + 
  \frac{\lambda}{2}\psi_\alpha^\dagger\psi_\beta^\dagger\psi_\beta\psi_\alpha\;,
  \label{lagrangian}
\end{equation}
where $\xi(\hatp) = \hatp^2/(2m) - \mu$. The chemical potentials for the two
species of fermions, $\psi_1$ and $\psi_2$, in terms of the average chemical
potential $\mu$ and the splitting $2\delta\mu$ are, $\mu_1=\mu+\delta\mu$ and 
$\mu_2=\mu-\delta\mu$, respectively. We are interested in the case where the
interaction between the two species of fermions is attractive, meaning
$\lambda>0$. In the BCS regime, then, for small enough $T$ and
$\delta\mu$, the phase of the system will be characterized by a non-zero
difermion condensate
\begin{equation}
\langle\psi_\alpha({\bf{r}})\psi_\beta({\bf{r}})\rangle =
\frac{1}{\lambda}\epsilon_{\alpha\beta}\Delta({\bf{r}})\label{difermion
condensate}\;.
\end{equation}
We are interested in the phase boundary between such a superfluid phase and the
normal phase, where $\Delta({\bf{r}})=0$.

In addition, the system will be characterized by specific profiles for the
number density of the two species\;,
\begin{equation}
\langle\psi_1^\dagger({\bf{r}})\psi_1({\bf{r}})\rangle =
n_1({\bf{r}})\;,\;\;\;\;\;
\langle\psi_2^\dagger({\bf{r}})\psi_2({\bf{r}})\rangle =
n_2({\bf{r}})\;.
\end{equation}
In our calculations we will find it more convenient to write $n_1$ and $n_2$ in
terms of the average density $n(r)=(1/2)(n_1({\bf{r}})+n_2({\bf{r}}))$ and the
difference in densities $\delta n(r)=(1/2)(n_1({\bf{r}})-n_2({\bf{r}}))$.

 To specify the system for any $T$, $\mu$, $\delta\mu$ for a given $\lambda$ (which
we will trade for the scattering length $a$), we need three equations that let
us solve for the three variables $\Delta$, $n$ and $\delta n$. These are the gap
and the number equations. For homogeneous condensates these take the form,
\begin{equation}
\frac{\partial \Omega}{\partial \Delta} = 0 \;,\;\;\;\;\;
\frac{\partial \Omega}{\partial \mu} = -2n \;,\;\;\;\;\;
\frac{\partial \Omega}{\partial \delta\mu} = -2\delta n \label{gap and number
equations defined}\;.
\end{equation}

 We calculate the free energy, $\Omega$, in a mean field approximation where we
replace the four Fermi interaction by its mean field value
\begin{equation}
\begin{split}
\frac{\lambda}{2}\psi_\alpha^\dagger({\bf{r}})\psi_\beta^\dagger({\bf{r}})\psi_\beta({\bf{r}})\psi_\alpha({\bf{r}})
\rightarrow
\ha\Delta^*({\bf{r}})\epsilon_{\alpha\beta}\psi_\alpha({\bf{r}})\psi_\beta({\bf{r}})
-\ha\Delta({\bf{r}})\epsilon_{\alpha\beta}\psi^\dagger_\alpha({\bf{r}})\psi^\dagger_\beta({\bf{r}})
-\frac{|\Delta(\bf{r})|^2}{\lambda}\\
+\lambda n_1({\bf{r}})\psi^\dagger_2({\bf{r}})\psi_2({\bf{r}})
+\lambda n_2({\bf{r}})\psi^\dagger_1({\bf{r}})\psi_1({\bf{r}})
-\lambda n_1({\bf{r}})n_2({\bf{r}})\;.
\end{split}
\end{equation}
The terms proportional to $n_1$ and $n_2$ give rise to Hartree corrections to
the free energy. 

 Upon making the mean field approximation, and performing standard manipulations, we can
write the lagrangian density (eq.~(\ref{lagrangian})) in a quadratic form in terms of the
Nambu-Gorkov spinor,
\begin{equation}
\Psi = \left(\psi_1\;\;
  \psi_2\;\;
  \psi_1^\dagger\;\;
  \psi_2^\dagger
   \right)^T\;.
\end{equation}
The final answer is
\begin{equation}
{\cal{L}} = \ha\Pd \left( 
\begin{array}{cc} 
i{\partial_t}-\txip+\tdm\sigma^3 & -\Delta(x)\varepsilon    \\
\Delta^*(x)\varepsilon &   i{\partial_t}+\txip-\tdm\sigma^3 
\end{array} \right)\Psi 
  + \delta({\bf 0})\tmu - \frac{|\Delta|^2}{\lambda}
  - \lambda(n^2(\bfr) - \dn^2(\bfr))
\label{lagrangian mean field}\;, 
\end{equation}
where $\txip = \hatp^2/(2m) - \tmu$. $\tmu$ and $\tdm$ include the Hartree terms
and are given by
\begin{equation}
\tmu = \mu + \lambda n \;,\;\;\;\;\;\;
\tdm = \dm - \lambda \dn\label{tmu}\;.
\end{equation}
That the Hartree corrections to the chemical potentials should be
this way can be seen immediately if we recall that the Hartree terms change
$\mu_1$ to $\tmu_1=\mu_1+\lambda n_2$ and $\mu_2$ to $\tmu_2=\mu_2 + \lambda
n_1$. The presence of the mysterious looking term $\delta({\bf 0})\tmu$ can be explained
as follows. 

 When written in terms of the fields $\psi$ and $\psi^\dagger$, the mean field
lagrangian density has a piece $\sum_\alpha \tmu_\alpha
\psi_\alpha^\dagger\psi_\alpha$. To write this in a symmetric form in terms of
the components of the Nambu-Gorkov spinor $\Psi$, we need to exchange the
ordering of $\psi$ and $\psi^\dagger$, which gives rise to the term in question.
More explicitly,
\begin{equation}
\sum_\alpha \tmu_\alpha
\psi_\alpha^\dagger\psi_\alpha = \ha\sum_\alpha(
 \tmu_\alpha
  \psi_\alpha^\dagger\psi_\alpha+
 \tmu_\alpha
  \psi_\alpha^\dagger\psi_\alpha) 
  = \ha\sum_\alpha(
 \tmu_\alpha
  \psi_\alpha^\dagger\psi_\alpha-
 \tmu_\alpha
  \psi_\alpha\psi_\alpha^\dagger+\tmu_\alpha\delta({\bf 0}))\;,
\end{equation}
where we have used the fermion anticommutation relation
\begin{equation}
\{\psi_\alpha(\bfr),\psi^\dagger_\beta(\bfr')\} =\delta_{\alpha\beta}\delta(\bfr-\bfr')
\;.
\end{equation}
This term is important in canceling out a divergent contribution to the free
energy, as we shall see below. A similar term occurs while reordering $\sum_\alpha 
\psi_\alpha^\dagger(i\partial_t - \hatp^2/(2m))\psi_\alpha $, but is the same in
normal and superfluid matter and does not affect the phase competition.

 In the mean field approximation the lagrangian density is bilinear in the
fermion fields and the free energy is found by direct integration over the
fields. We find 
\begin{equation}
\begin{split}
\intspace{x_E} \Omega &= \intspace{x_E}\Bigl\{ 
  - \delta({\bf 0})\tmu + \frac{|\Delta(\bfr)|^2}{\lambda}
  + \lambda(n^2(\bfr) - \dn^2(\bfr))\Bigr\} \\
&\phantom{++}- \Bigl\{\bigl[\frac{1}{2}\tr\log \left( 
\begin{array}{cc} 
{-\partial_{x^4}}-\txip+\tdm & -\Delta(\bfr)    \\
-\Delta^*(\bfr)  &   {-\partial_{x^4}}+\txip+\tdm
\end{array} \right)+(\tdm\rightarrow -\tdm)\bigr]\Bigr\}\label{Omega1}\;,
\end{split}
\end{equation}
where $x_E$ represents the euclidean space four vector $(x^4,\bfr)$, with $\bfr$
lying in position space with volume $V$ and $x^4\in[-1/(2T),1/(2T)]$.

  We now compare the free energies of the superfluid phases to the normal phase to
find where the boundary between the phases lies.

\section{Phase boundary between the superfluid and normal
phase~\label{section:phase boundary}}
 
 We first consider the competition between the homogeneous superfluid phase and
the normal phase, and then the competition between inhomogeneous superfluids and
the normal phase.
\vspace{-0.2in}
\subsection{Homogeneous superfluid and the normal phase\label{sec:homogeneous}}

 The homogeneous phases are defined by the condition that $\Delta$, $n$ and
$\dn$ are all independent of $\bfr$. The argument of the $\log$ in
eq.~(\ref{Omega1}) is then diagonal in momentum space and the free energy
density is simply, 
\begin{equation}
\begin{split}
\Omega(\Delta,n,\dn,T,\mu,\dmu) &= \Bigl\{ 
  - \tmu\threeier{\bfp}(1) + \frac{|\Delta|^2}{\lambda}
  + \lambda(n^2 - \dn^2)\Bigr\} \\
&\phantom{++}- \Bigl\{\threeier{\bfp} \bigl[\frac{T}{2}\sum_{{p^4=}\atop{(2n+1)\pi T}}\log \Bigl( 
(ip^4+\tdm-\tepsp)(ip^4+\tdm+\tepsp)
\Bigr)+(\tdm\rightarrow -\tdm)\bigr]\Bigr\}\\
 &= \Bigl\{  \frac{|\Delta|^2}{\lambda}
  + \lambda(n^2(\bfr) - \dn^2(\bfr))\Bigr\} \\
&\phantom{++}-\threeier{\bfp} \Bigl\{T\Bigl[ 
  \log \bigl[\cosh\Bigl(\frac{\tdmu+\tepsp}{2T}\Bigr)\bigr]
 +\log \bigl[\cosh\Bigl(\frac{-\tdmu+\tepsp}{2T}\Bigr)\bigr]\Bigr]
   + \tmu\Bigr\}\label{OmegaHomogenous}\;,
\end{split}
\end{equation}
where $\tepsp=\surd(\tilde{\xi}^2(\bfp)+|\Delta|^2)$ and we have rewritten
$\delta({\bf{0}})$ as an intergal over momentum space. 

The gap equation is the condition that the free energy is stationary with
respect to small variations in the magnitude of $\Delta$. (A position
independent phase of $\Delta$ does not affect the free energy and hence, for
simplicity, we will take $\Delta$ to be real and positive for the rest of the
section~(\ref{sec:homogeneous}).)
\begin{equation}
\begin{split}
0 =& \frac{\partial{\Omega}}{\partial{\Delta}} \\
  =& \frac{2\Delta}{\lambda} - \ha\threeier{\bfp} \frac{\Delta}{\tepsp}\Bigl[ 
  \tanh\Bigl(\frac{\tdmu+\tepsp}{2T}\Bigr)
 +\tanh\Bigl(\frac{-\tdmu+\tepsp}{2T}\Bigr)\Bigr]\label{gap equation}\;.
\end{split}
\end{equation}
The trivial solution, $\Delta=0$, corresponds to the normal phase. A competing
superfluid phase exists if eq.~(\ref{gap equation}) has a solution with
$\Delta\neq 0$.

 As it stands, the second term in eq.~(\ref{gap equation}) is linearly divergent
in $\bfp$ and needs to be regulated in some manner. One way is to cut off the
momentum integration at some momentum $\Lambda$, chosen to be sufficiently
larger than the Fermi momentum, $k_F=\surd(2m\mu)$, to capture all the features of the intergand. In the weak
coupling (BCS) regime, it enough to take $\Lambda$ to be several times $k_F$. The solution of
eq.~(\ref{gap equation}) then depends on $\lambda$ and $\Lambda$, but it is
useful to rewrite the results in terms of a physical observable. A popular
choice is to use the relation between the $s$-wave scattering length $a$, and
$\lambda$.
\begin{equation}
\frac{1}{\lambda}  = \frac{-m}{4\pi a} +
\threeier{\bfp}\frac{m}{\bfp^2}\label{scattering length}\;.
\end{equation}
Canceling $2\Delta\neq 0$ from the right hand side of the gap equation,
eq.~(\ref{gap equation}), and substituting $1/\lambda$ in
the first term in eq.~(\ref{gap equation}) from eq.~(\ref{scattering length}),
we obtain the relation
\begin{equation}
-\frac{m}{4\pi a} = \threeier{\bfp}\Bigl\{\frac{1}{4\tepsp}\Bigl[ 
  \tanh\Bigl(\frac{\tdmu+\tepsp}{2T}\Bigr)
 +\tanh\Bigl(\frac{-\tdmu+\tepsp}{2T}\Bigr)\Bigr] - \frac{m}{\bfp^2}
 \Bigr\}\label{gap equation2}\;,
\end{equation}
which is now cutoff independent. Recall, however, that $\lambda$ also appears
implicitly in the definitions of $\tmu$ and $\tdmu$~(eq.~(\ref{tmu})) where it
multiplies $n$ and $\delta n$, respectively. In all these places, we replace
$\lambda$ by $-(4\pi a)/m$. To see why this is reasonable, consider
eq.~(\ref{scattering length}) with a momentum cut off
$\Lambda$.
\begin{equation}
\frac{1}{\lambda} = -\frac{m}{4\pi a} + m\Lambda\label{scattering length cutoff}\;.
\end{equation}
If $|4\pi\Lambda a|\ll 1$, then in eq.~(\ref{scattering length cutoff}) we can ignore 
the second term on the right hand side compared to the first term. In the weak
coupling limit $|\pi k_F a|\ll 1$. If $\Lambda$ is taken
to be not many times larger than $k_F$, as we argued can be done in the weak
coupling regime, then indeed we can take $\lambda$ to be $-(4\pi a)/m$ to a good
approximation.

 To summarize, the gap equation is given by eq.~(\ref{gap equation2}) with
\begin{equation}
\tmu = \mu + \Bigl(\frac{-4\pi a}{m}\Bigr) n\;,\;\;\;\;
\tdmu = \dmu - \Bigl(\frac{-4\pi a}{m}\Bigr) \dn\label{tmu approx}\;.
\end{equation}

 The number equations are found by explicitly calculating the derivatives with
respect to $\mu$ and $\dmu$~(eq.~(\ref{gap and number equations defined})). The
final expressions are given below.
\begin{equation}
-n = \frac{1}{4} \threeier{\bfp}\Bigl\{\Bigl[\tanh\Bigl(\frac{\tdmu+\tepsp}{2T}\Bigr)
 +\tanh\Bigl(\frac{-\tdmu+\tepsp}{2T}\Bigr)\Bigr] \frac{\txip}{\tepsp} -
 2\Bigr\}\label{n equation}\;,
\end{equation}
and,
\begin{equation}
-\dn = \frac{1}{4} \threeier{\bfp}\Bigl\{\Bigl[-\tanh\Bigl(\frac{\tdmu+\tepsp}{2T}\Bigr)
 +\tanh\Bigl(\frac{-\tdmu+\tepsp}{2T}\Bigr)\Bigr]\Bigr\}\label{dn equation}\;,
\end{equation}
with $\tmu$ and $\tdmu$ given by eq.~(\ref{tmu approx}).

 For the normal phase, $\Delta=0$, and we can solve eqs.~(\ref{n
equation},\ref{dn equation}) to find $n$ and $\dn$ to obtain values we will call $n_N$ and
$\dn_N$ respectively. The free energy of the normal phase is given by,
\begin{equation}
\begin{split}
\Omega_N(\mu,\dmu,T)  
 &= \Bigl\{  \Bigl(\frac{-4\pi a}{m}\Bigr)(n_N^2 - \dn_N^2)\Bigr\} \\
&\phantom{++}-\threeier{\bfp} \Bigl\{T\Bigl[ 
  \log \bigl[\cosh\Bigl(\frac{\tdmu_N+\txip_N}{2T}\Bigr)\bigr]
 +\log \bigl[\cosh\Bigl(\frac{-\tdmu_N+\txip_N}{2T}\Bigr)\bigr]\Bigr]
   + \tmu_N\Bigr\}\label{OmegaNormal}
\end{split}
\end{equation}
with $\txip_N=\hat{\bfp}^2/(2m)-\tmu_N$ and
\begin{equation}
\tmu_N = \mu + \Bigl(\frac{-4\pi a}{m}\Bigr) n_N\;,\;\;\;\;\;
\tdmu_N = \dmu - \Bigl(\frac{-4\pi a}{m}\Bigr) \dn_N\label{tmuN approx}\;.
\end{equation}

\begin{center}
\begin{figure}[h!]
\includegraphics[width=4.5in,angle=0]{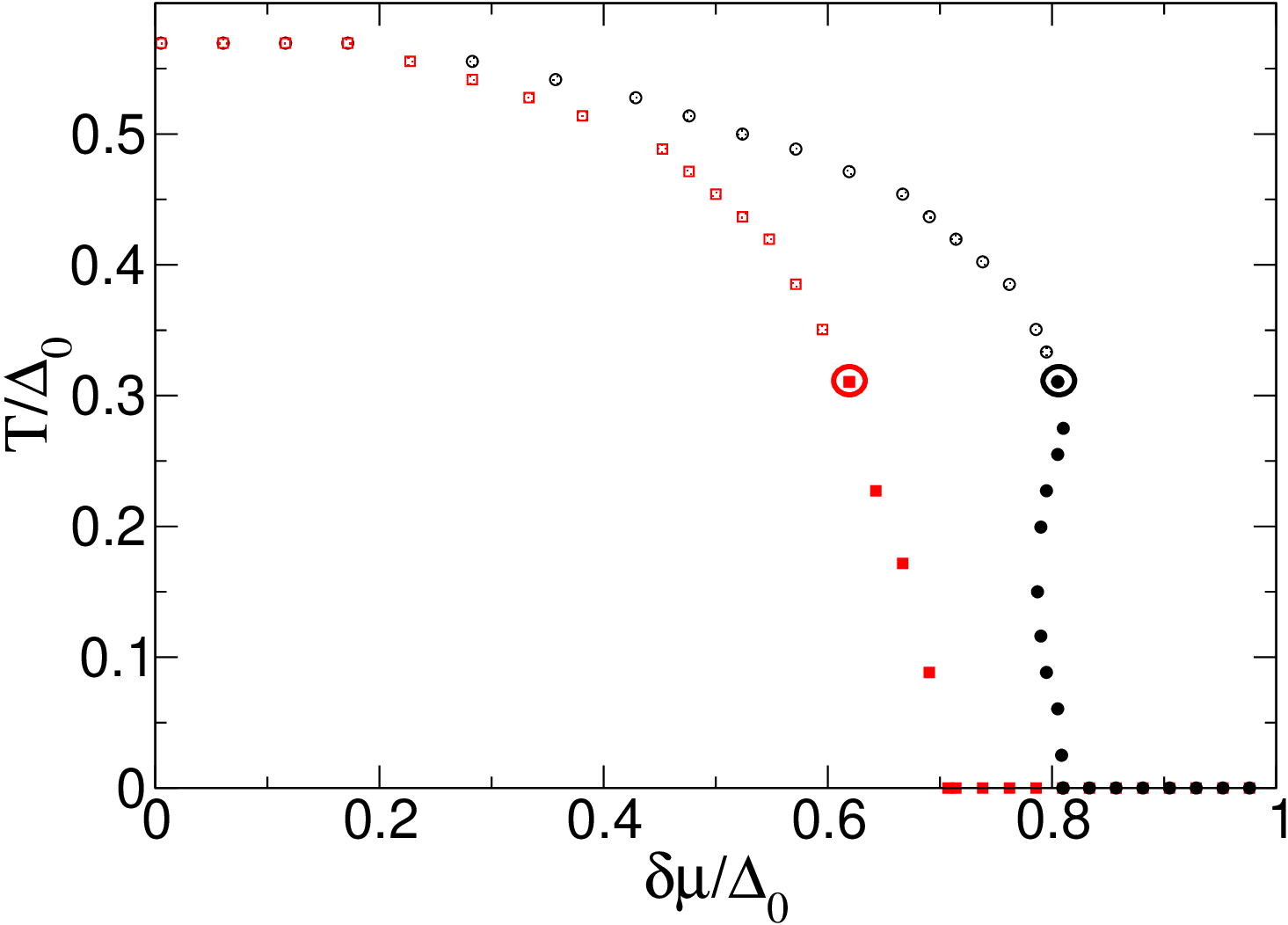}
\caption{(colored online) We show the phase boundary between normal and homogeneous superfluid
phases in $\dm$, $T$ space, for $g=-0.72$. In the
curve marked by circles (black online), we include the Hartree corrections. The value 
of $\Delta_0/\mu$ is $0.058$.  The curve marked by
squares (red online) does not include Hartree corrections. The value of 
$\Delta_0/\mu$ in this case is $0.03$. At $T=0$,
the first order phase transition from the superfluid to the normal phase
occurs at $\dmu/\Delta_0\sim 0.71$, consistent with weak coupling results. 
The filled points mark the region where the phase transition is first order
while the hollow points mark a second order transition. The points
where the transition changes from being second order to becoming first order have been
encircled.} 
\label{homogeneous-normal}
\end{figure}
\end{center}

 If, in addition, eqs.~(\ref{gap equation2},\ref{n equation},\ref{dn equation}) possess
solutions with $\Delta\neq 0$, $n_\Delta$ and $\dn_\Delta$, we need to compare the 
free energies of these superfluid solutions to $\Omega_N$. The difference in the 
free energies is given by
\begin{equation}
\begin{split}
(\Omega_s-\Omega_N)(\mu,\dmu,T)
 &= \Bigl\{  -\frac{m\Delta^2}{4\pi a}
  + \Bigl(\frac{-4\pi a}{m}\Bigr)(n_\Delta^2 -n_N^2- \dn_\Delta^2+\dn_N^2)\Bigr\} \\
&\phantom{++}-\threeier{\bfp} \Bigl\{T\Bigl[ 
  \log \bigl[\cosh\Bigl(\frac{\tdmu+\tepsp}{2T}\Bigr)\bigr]
 +\log \bigl[\cosh\Bigl(\frac{-\tdmu+\tepsp}{2T}\Bigr)\bigr]\Bigr]
   + \tmu-\frac{m\Delta^2}{\bfp^2}\Bigr\}\label{OmegaDiff}\;.
\end{split}
\end{equation}
If $\Omega_s-\Omega_N>0$ then the normal phase is favored over the superfluid
phase, and vice versa. In Fig.~(\ref{homogeneous-normal}) we look at the boundary
marking the normal to superfluid phase transition in $T$, $\dm$ space for
$g=(\pi k_F a)^{(-1)} = -0.72$. For different values of $\mu$ and $m$, if we
choose $a$ so that the dimensionless parameter $g$ remains the same, then the
physical quantities scale with $\mu$ and $m$ as follows.
\begin{equation}
\begin{split}
n_\alpha(m,\mu,T,\dm) &=  m^{(3/2)}\mu^{(3/2)}n_\alpha(1,1,(T/\mu),(\dmu/\mu))\\
\Delta(m,\mu,T,\dm) &=  \mu\Delta(1,1,(T/\mu),(\dmu/\mu))\\
\Omega(m,\mu,T,\dm) &= m^{(3/2)}\mu^{(5/2)}\Omega(1,1,(T/\mu),(\dmu/\mu))\;.
\end{split}
\end{equation}
In particular, the value of $\Delta_0$, the value of the gap parameter at $T=0$
and $\dm=0$, scales linearly with $\mu$. To remove the dependence of the phase
transition curve on the overall scales, it is useful to draw the phase diagram in
terms of dimensionless variables. Since $\Delta_0$ is proportional to $\mu$, we
take the $x-axis$ to be $\dmu/\Delta_0$ and the $y-axis$ to be $T/\Delta_0$. 

 For comparison, in Fig.~(\ref{homogeneous-normal}) we also show the result
when one does not include the Hartree corrections. One effect of 
Hartree corrections is to simply shift the chemical potentials. For example, at
$T=0$ and $\delta\mu=0$, including these increases the
``effective'' chemical potential, $\tmu=\mu-(4\pi a/m)n$. (Recall that $a<0$ in the BCS
regime.) Therefore, for the same $a$, the value of $\Delta_0$ is greater when we include the Hartree
corrections, compared to when we do not include them. To get rid of this overall
change in $\Delta_0$, we scale $T$ and $\delta\mu$ in the ``non-Hartree'' curve 
by the ``non-Hartree'' value of $\Delta_0$. The two curves are still 
different, and that has to do with the fact that including Hartree corrections
affects the competition between the normal and the superfluid phases. Number densities, and 
therefore Hartree corrections, are different in the two phases at the 
phase boundary, if the phase transition between the two phases is first order.
In the following paragraphs, we discuss the effect of Hartree corrections on the phase transition curve
between the normal and the superfluid phase in more detail.

First looking at $T=0$, $\delta\mu/\Delta_0$ at the first order transition is
larger than the weak coupling value $0.707$, as pointed out earlier
in~\cite{Carlson:2005}. Since at $T=0$, $\dn=0$ in the superfluid phase (the
superfluid phase is gapped), this is not simply due to a reduction in the
effective splitting between the Fermi surfaces in the superfluid region ($\tdmu
= \dmu \sim 0.81\Delta_0$ for the superfluid phase at the phase transition). As
discussed above, the change is due to the fact that $n$ and $\dn$ change
abruptly at the first order phase boundary. More specifically, there are two
effects both of which drive the phase transition to larger $\dmu$. Firstly,
$\dn$ is positive in the normal phase, implying $\tdmu$ is smaller than $\dmu$,
which increases the free energy of the normal phase. Secondly, $n$ is larger in
the superfluid phase because of pairing, and this also makes the superfluid
phase more favorable.
 
 Next, looking along the $y-axis$, at $\dmu=0$, the transition from the
superfluid to the normal phase at the critical temperature is second order, and
the number densities in the two phases are the same at the critical
temperature.  It is understandable, therefore, that the transition occurs at
$T_c/\Delta_0\sim 0.567$, the standard weak coupling value.
  
\begin{center}
\begin{figure}[h!]
\includegraphics[width=4.5in,angle=0]{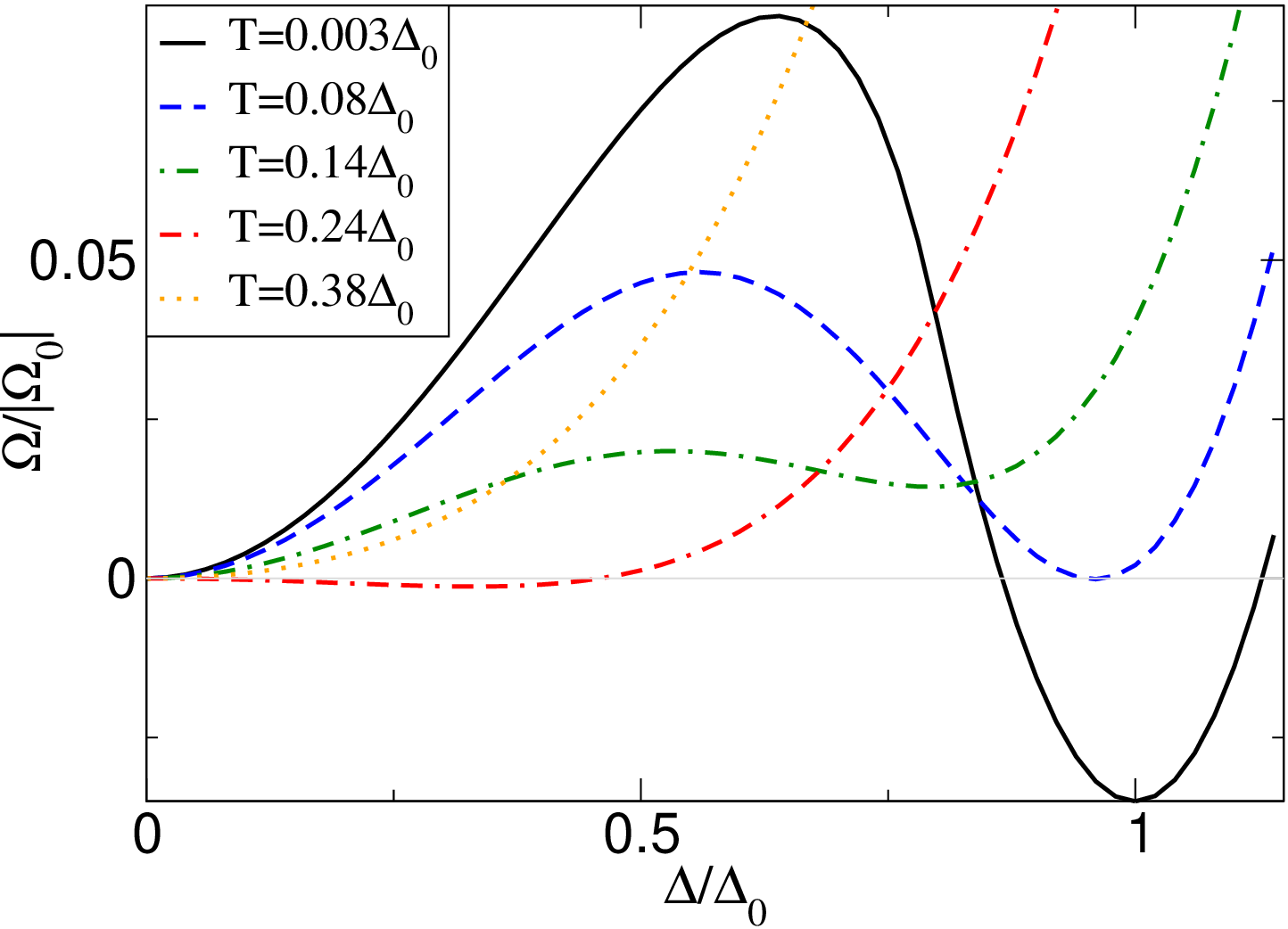}
\caption{(colored online) This figure shows the free energy as a function of $\Delta$ at
$\dmu/\Delta_0 = 0.795$ for various values of $T$. The solid curve (black
online) corresponds to
the lowest temperature, $T/\Delta_0=0.003$, and clearly shows a robust
superfluid phase with $\Delta\sim\Delta_0$. The dashed curve (blue online)  
corresponds to
$T/\Delta_0=0.08$ and is the point of a first order phase transition to the
normal phase. A ``fragile'' superfluid phase reappears 
before it disappears again at a higher temperature.} 
\label{OmegaDeltaplot}
\end{figure}
\end{center}

  Things are interesting close to $\dmu/\Delta_0\sim0.8$, where the shape of the
curve is qualitatively altered. For a window of splittings,
$\dmu/\Delta_0\in(0.79,0.81)$, as we increase the temperature, we encounter not
one but three normal-superfluid transitions. To clarify how this comes about,
consider the shape of $\Omega$ as a function of $\Delta$ for various values of
the temperature, at $\dmu/\Delta_0= 0.795$~(Fig.~\ref{OmegaDeltaplot}). At $T=0$, the
local minimum at $\Delta=\Delta_0$ is favored. As we increase the temperature,
this minimum becomes shallower and eventually there is a first order transition
to the normal phase. As we keep increasing the temperature, the $\Delta=0$
solution becomes unstable and the superfluid state is favored again for a range
of temperatures. Eventually, at a large enough temperature, there is a second
order transition to the normal phase. This reappearance of superfluidity at
higher temperatures can be understood intuitively as follows. At zero
temperature, BCS pairing is stressed due to a non-zero $\dm$, because fermions
of the two species can not find partners of opposite momenta lying on the
distinct Fermi surfaces determined by their different chemical potentials. BCS pairing 
in such systems requires the two distinct Fermi spheres to equalize at a Fermi
momentum different from the value given by the corresponding chemical
potentials, costing free energy. At non-zero temperatures, however, the Fermi
surfaces are smeared and it is possible to find partners of opposite momenta
even without equalizing the Fermi surfaces. Such an effect of temperature on
pairing has been noted previously in the context of pairing between $u$ and $d$
quarks in the 2SC
phase~\cite{shovkovy-2003-564,huang-2003-729,fukushima-2005-71,blaschke-2005-72,abuki-2006-768,
ruester-2006-73,warringa-2006}, and a nuclear physics
context~\cite{sedrakian-2000-84}. 

 To see in a different way why there are three transitions, it is useful to expand  
the free energy in powers of $\Delta^2$ (Ginzburg-Landau expansion) and look at
how the coefficient of the quadratic term changes as we change $T$, keeping
$\dm$ constant. We write the free energy as
$\alpha\Delta^2+{\cal{O}}(\Delta^4)$. $\alpha<0$ points to an instability in the
$\Delta=0$ state to the formation of a non-zero condensate, while $\alpha>0$
means that the normal state is locally stable, but does not tell us whether it is
globally favored or not. 

\begin{center}
\begin{figure}[h!]
\includegraphics[width=4.5in,angle=0]{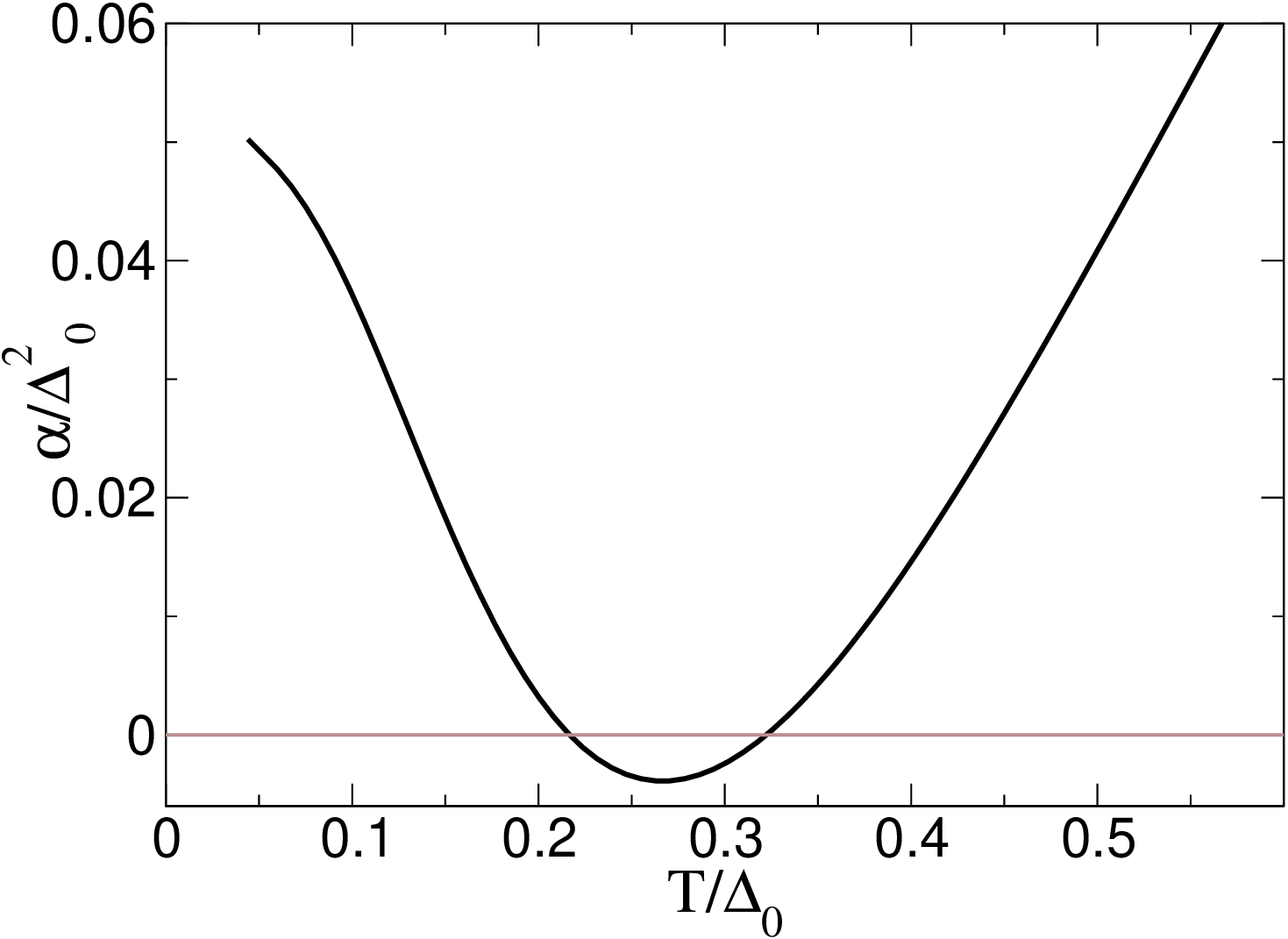}
\caption{Plot of $\alpha$ as a function of $T/\Delta_0$ for
$\dmu=0.795\Delta_0$. The normal phase is locally stable for $T/\Delta_0\in
(0,0.22)$, locally unstable for $T/\Delta_0\in(0.22,0.33)$, and locally stable
for $T$ larger than $0.33\Delta_0$.} 
\label{alphaplot}
\end{figure}
\end{center}

 In Fig.~(\ref{alphaplot}) we plot $\alpha$ (in units of $\Delta_0^2$) as a function of
$T/\Delta_0$, for $\dmu=0.795\Delta_0$ kept constant. For small $T$, the normal
phase is locally stable, but globally disfavored to the $\Delta=\Delta_0$
state~(Fig.~(\ref{OmegaDeltaplot})). As we increase $T$, the superfluid phase
becomes less and less favorable until finally at $T/\Delta_0=0.08$ the normal
phase is globally favored. On increasing $T$ further, we find that the normal
phase becomes locally unstable in the region $T/\Delta_0\in(0.22,0.33)$, and
this gives rise to a region of ``fragile'' superfluidity.

 For smaller couplings the window of $\dm/\Delta_0$ where 
we encounter three normal-superfluid transitions, becomes narrower, 
while for stronger couplings it becomes wider.

\subsection{Inhomogeneous condensates\label{sec:inhomogeneous}}

 Let us now consider the case where $\Delta$, $n$ and $\dn$ depend on
$\bfr$. The argument of the $\log$ in eq.~(\ref{Omega1}) is no longer diagonal 
in momentum space and hence we evaluate the free energy in a Ginzburg-Landau
expansion. We are interested in finding the curve along which the normal phase becomes unstable
to the growth of an inhomogeneous condensate. Working in the limit of small
$\Delta(\bfr)$, we expand the $\log$ in $\Delta(\bfr)$ and drop terms
proportional to $\Delta^4$ and higher, and obtain
\begin{equation}
\begin{split}
\int d^3\bfr \Omega &= \int d^3\bfr \Bigl\{
  - \tmu\threeier{\bfp}(1) + \frac{|\Delta(\bfr)|^2}{\lambda}
  + \lambda(n^2 - \dn^2)\Bigr\} \\
&\phantom{++}- \Bigl\{T\sum_{{p^4=}\atop{(2n+1)\pi T}}\bigl[
{\rm{Tr}}_{\bfr}\log \bigl((ip^4+\tdm-\txip)(ip^4+\tdm+\txip)\bigr)\\
&\phantom{+++++++}-{\rm{Tr}}_{\bfr}\bigl((ip^4+\tdm-\txip)^{(-1)}
  \Delta(\bfr)(ip^4+\tdm+\txip)^{(-1)}\Delta^*(\bfr)\bigr)
\bigr]\Bigr\}
\end{split}
\end{equation}
We now argue that within the approximations we are working in, we can replace
$n$ and $\dn$ by their values in normal matter to compute $\Omega$. It is
obvious that we can do so in the term,
\begin{equation}
T\sum_{{p^4=}\atop{(2n+1)\pi T}}
\bigl((ip^4+\tdm-\txip)^{(-1)}\Delta(\bfr)(ip^4+\tdm+\txip)^{(-1)}\Delta^*(\bfr)\bigr)
\label{Omega term1}\;,
\end{equation}
because the corrections to $n$ and $\dn$ due to pairing is proportional to
$\Delta^2$, and keeping these corrections in eq.~(\ref{Omega term1}) (where they
appear in $\tmu$ and $\tdmu$) will only change the result by order $\Delta^4$.
There are ${\cal{O}}(\Delta^2)$ contributions to $\Omega$ from
$\tmu\threeier{\bfp}(1)$,  $\lambda(n^2 - \dn^2)$ and 
${\rm{Tr}}_{\bfr}\log \bigl((ip^4+\tdm-\txip)(ip^4+\tdm+\txip)\bigr)$, but in
all these cases, the $\Delta^2$ correction to $n$ or $\dn$ is further
multiplied by $\lambda$ and for weak coupling, this should give a small overall
contribution. (This is the $\Delta^2$ correction to the Hartree term which
itself is a correction). In the approximation where we neglect this contribution, 
when we calculate the difference between the
normal and superfluid free energies, these three terms cancel out and we obtain
\begin{equation}
\begin{split}
\int d^3\bfr ( \Omega(\Delta(\bfr))-\Omega_N ) =&  \int d^3\bfr \Bigl\{
   \frac{|\Delta(\bfr)|^2}{\lambda} \Bigr\}\\
&\phantom{+}+\threeier{\bfp}\Bigl\{T\sum_{{p^4=}\atop{(2n+1)\pi T}}\bigl[
{\rm{Tr}}_{\bfr}\bigl((ip^4+\tdm_N-\txip_N)^{(-1)}\Delta(\bfr)
(ip^4+\tdm_N+\txip_N)^{(-1)}\Delta^*(\bfr)\bigr)\bigr]\Bigr\}\;.
\end{split}
\end{equation}
In momentum space this gives, 
\begin{equation}
\begin{split}
\int d^3\bfr ( \Omega(\Delta(\bfr))-\Omega_N ) = &\threeier{\bfk}
\Delta(\bfk)\Delta^*(-\bfk)\Bigl[\frac{-m}{4\pi a}+\\
&+\threeier{\bfp} \Bigl\{\frac{m}{\bfp^2}+T\sum_{{p^4=}\atop{(2n+1)\pi
T}}(ip^4+\tdm_N-\tilde{\xi}(\bfp+\bfk)_N)^{(-1)}
(ip^4+\tdm_N+\tilde{\xi}(\bfp)_N)^{(-1)}\Bigr\}\Bigr]\;,
\end{split}
\end{equation}
where we have used eq.~(\ref{scattering length}) to rewrite $\lambda$.

 It is convenient at this point, to separate the ``potential'' contribution (the
contribution independent of $\bfk$) from the ``gradient'' contribution (zero for
$\bfk={\bf{0}}$)~\cite{gubankova-2008}. We write 
\begin{equation}
\int d^3\bfr ( \Omega(\Delta(\bfr))-\Omega_N ) = \threeier{\bfk}
\Delta(\bfk)\Delta^*(-\bfk)\Bigl\{   \alpha+f(|\bfk|) \Bigr\} \label{GL about
zero}
\end{equation}
with,
\begin{equation}
\begin{split}
\alpha  &= -\frac{m}{4\pi a} + \threeier{\bfp} \Bigl\{\frac{m}{\bfp^2} + 
  T\sum_{{p^4=}\atop{(2n+1)\pi T}}(ip^4+\tdm_N-\tilde{\xi}(\bfp)_N)^{(-1)}
(ip^4+\tdm_N+\tilde{\xi}(\bfp)_N)^{(-1)}\Bigr\}\\
  &=-\frac{m}{4\pi a} + \threeier{\bfp} \Bigl\{\frac{m}{\bfp^2} 
  - \Bigl[\tanh\bigl(\frac{\tdmu_N+\txip_N}{2T}\bigr)
  +\tanh\bigl(\frac{-\tdmu_N+\txip_N}{2T}\bigr)\Bigr]\frac{1}{4\xip} \Bigr\}
\end{split}
\end{equation}
and,
\begin{equation}
\begin{split}
f(|\bfk|) &=\frac{1}{2}\threeier{\bfp} 
\Bigl\{T\sum_{{p^4=}\atop{(2n+1)\pi
T}}\frac{(\tilde{\xi}(\bfp+\bfk)_N-\tilde{\xi}(\bfp)_N)^2}{((ip^4+\tdm_N)^2-(\tilde{\xi}(\bfp)_N)^2)
((ip^4+\tdm_N)^2-(\tilde{\xi}(\bfp+\bfk)_N)^2)}\Bigr\}\\
  &=\frac{1}{2}\threeier{\bfp} 
\Bigr\{\frac{(\tilde{\xi}(\bfp+\bfk)_N-\tilde{\xi}(\bfp)_N)}
{(\tilde{\xi}(\bfp+\bfk)_N+\tilde{\xi}(\bfp)_N)}
\Bigl[\frac{g(\tilde{\xi}(\bfp)_N)}{\tilde{\xi}(\bfp)_N}
-\frac{g(\tilde{\xi}(\bfp+\bfk)_N)}{\tilde{\xi}(\bfp+\bfk)_N}\Bigr]\Bigr\}\;,
\end{split}
\end{equation}
where
\begin{equation}
g(\xi)=\frac{1}{2}\Bigl[\tanh\bigl(\frac{\tdmu+\xi}{2T}\bigr)
+\tanh\bigl(\frac{-\tdmu+\xi}{2T}\bigr)\Bigr]\;.
\end{equation}

\begin{center}
\begin{figure}[h!]
\includegraphics[width=4.5in,angle=0]{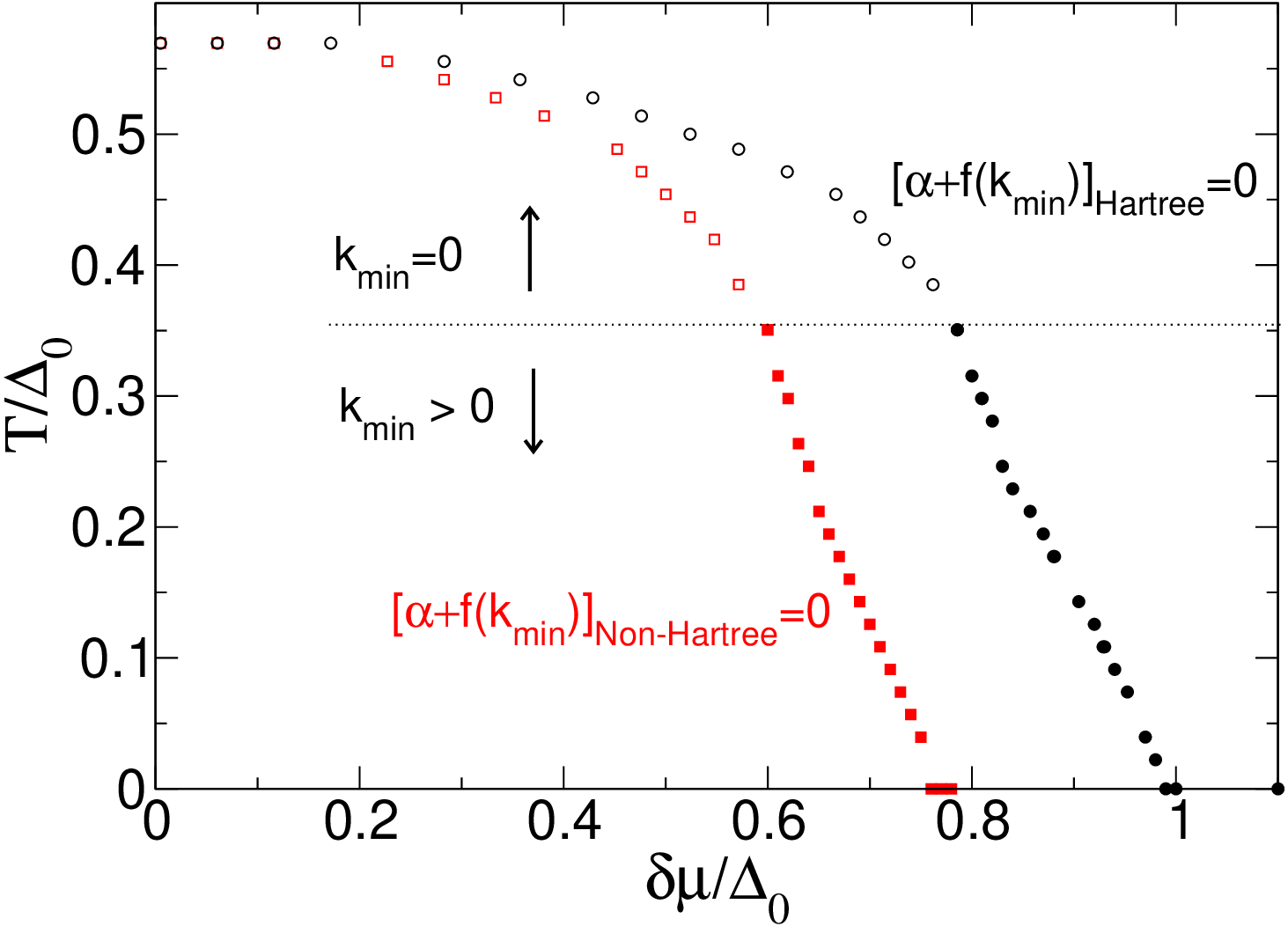}
\caption{(colored online) We show the curve marking a second order
phase transition between normal and inhomogeneous superfluid
phases in $\dm$, $T$ space, for $g=-0.72$. In the
curve marked by circles (black online), we include the Hartree corrections while
the curve marked by squares (red online) does not include these.
Where the symbols are hollow, the value of $k$ at which $\alpha+f(k)$
becomes zero, is zero, meaning that the instability is towards the formation of
homogeneous condensates. It is reasonable that at smaller $\dm$ and larger
temperatures, the formation of homogeneous condensates is favored because the
Fermi surface is smeared out and it is no longer advantageous to form Cooper
pairs with non-zero
net momenta.} 
\label{loff-normal}
\end{figure}
\end{center}

 At any given temperature, for large enough $\dmu$, the normal phase will be
favored over a phase with non-zero $\Delta(\bfk)$, and the combination
$\alpha+f(k)$ will be positive for all values of $k=|\bfk|$. As we decrease
$\delta\mu$ keeping $T$ constant, $\alpha+f(k)$ may become zero, and then negative, for a single mode
with momentum $k=k_{\rm{min}}$. If $k_{\rm{min}}\neq{{0}}$, this point symbolizes the
onset of the instability towards the formation of an inhomogeneous condensate.
At lower $\dmu$, more momentum modes may become unstable. If the transition from 
normal to inhomogeneous superfluidity is actually first
order, then we expect it to occur for values of $\delta\mu$ larger than the
value we find using this second order analysis.

 Fig.~(\ref{loff-normal}) shows the curve in $T$, $\dm$ space, which tells us the
value of $\dmu$ where the coefficient $\alpha+f(k)$ becomes zero for some $k$,
as we decrease $\dmu$ from a large value keeping $T$ constant. Since we are
looking at a second order phase transition line, once we take into account the
shift in the chemical potentials due to the Hartree corrections, the results are
consistent with the well known results for weak coupling. For example, at $T=0$,
the value of $\tdmu = \dmu-(-4\pi a/m)\dn$ at the phase transition is given by
$\tdmu/\Delta_0\sim 0.75$. Weak coupling Ginzburg-Landau calculations tell us
that the transition should occur at $\dmu^* = 0.754\Delta_0$ with the most
unstable momentum $k_{\rm{min}}$ given by $v_F k_{\rm{min}}/2 = 1.2 \dmu^*$. We
find that the value of $k$ which becomes unstable at $\tdmu/\Delta_0\sim 0.75 $
satisfies $v_F k_{\rm{min}}/2 \sim 1.24\tdmu $. For reference we also show the
LOFF boundary excluding the Hartree corrections using square dots (red online). The instability
towards inhomogeneous condensates at $T=0$ develops at $\dm/\Delta_0 \sim 0.75$,
and the value of $v_F k_{\rm{min}}/2$ we find at this point is $1.37\delta\mu$.

\section{Conclusions\label{section:conclusions}}

\begin{center}
\begin{figure}[h!]
\includegraphics[width=4.5in,angle=0]{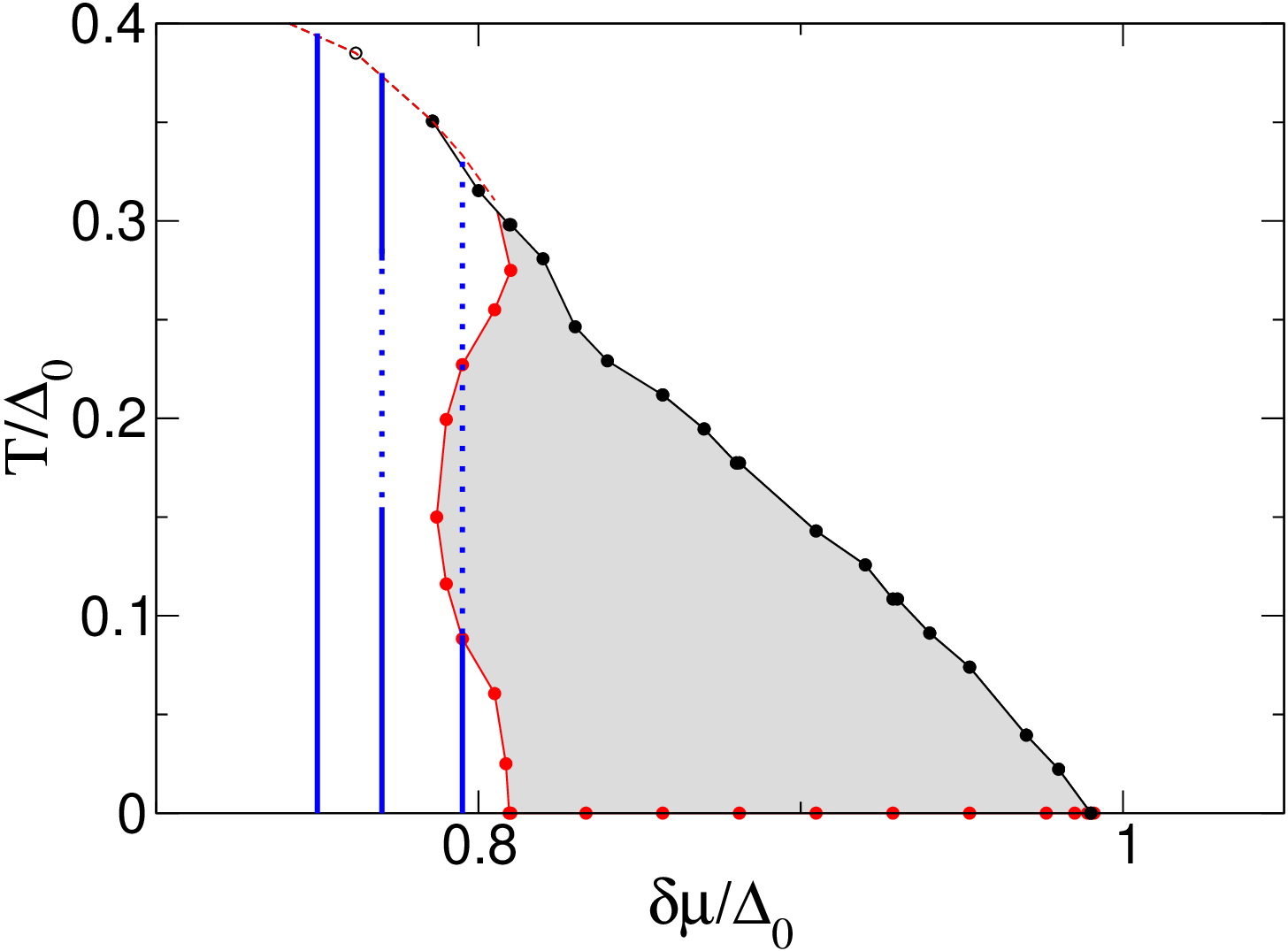}
\caption{(colored online) We show the phase boundaries between the normal and
the inhomogeneous superfluid phases and the normal and the homogeneous superfluid
phase, for $g=-0.72$. The lower curve (red online) represents the homogeneous
superfluid boundary, while the upper curve (black online) marks the instability
towards inhomogeneity. The two curves coincide for $\dm\lesssim 0.77\Delta_0$. The
shaded region is where LOFF-like phases may exist.  The three parallel thick
lines (blue online) placed at $\dm/\Delta_0=0.750$, $0.770$ and $0.795$, show
the stability of the mean field background solution to the growth of
inhomogeneities. In the temperature range for which a line is solid, the
homogeneous condensate is stable, while for the temperatures for which a line is
dotted, the condensate is unstable. The input for these lines comes from the
data shown in Fig.~(\ref{instability}).} 
\label{loff-pt}
\end{figure}
\end{center}

  We calculate the effect of Hartree corrections on the phase transition curve
separating the normal from the superfluid phase in the $T$, $\dm$ plane, in the
BCS regime. Our analysis is similar in spirit to the calculations performed by
the authors of~\cite{gubbels-2006-97}, who however concentrate on the unitary
regime. In their work, the Hartree corrections are taken to be of a well
motivated form~\cite{fregoso-2006-73} chosen so as to reproduce the data from
Monte Carlo simulations performed at unitarity~\cite{Carlson:2005}. We work in
the BCS regime, where the Hartree corrections can be calculated from first
principles, as described above.  Furthermore, we consider the instability of the
homogeneous states with respect to the formation of inhomogeneous condensates.

  The results are shown in Figs.~(\ref{homogeneous-normal},~\ref{loff-normal}) for
$g=-0.72$. In figure Fig.~(\ref{loff-pt}), we zoom in on the large $\delta\mu$
region and show the homogeneous-normal and inhomogeneous-normal phase boundaries
on the same diagram. The curve that is non-zero up to
$\delta\mu\sim0.99\Delta_0$ (black online) shows the parameter values for which the
coefficient of the $\Delta({\bf{k}})^2$ term, in a Ginzburg-Landau
expansion about the normal phase, becomes negative for some value of
$k=k_{\rm{min}}$.  This signals an instability towards the formation of an
inhomogeneous condensate. The curve that meets the $x-$axis at
$\delta\mu\sim0.81\Delta_0$ (red online) represents the points where the free energies of the
homogeneous superfluid and the normal phase are equal. Therefore, the region
between the boundaries, highlighted by a shading (gray online), is the region
where we expect LOFF-like phases to exist. Our first conclusion is that at weak
coupling, Hartree corrections do not destroy the parameter space where LOFF-like
phases may exist.

 Secondly, we find that the shape of the curve marking the transition from the
homogeneous superfluid phase to the normal phase, is qualitatively
altered~(Fig.~(\ref{homogeneous-normal})). For a narrow range of $\dmu$ near the
first order transition, we encounter three normal-superfluid transitions as we
increase the temperature keeping $\dmu$ constant. There is a first order phase
transition from the superfluid to the normal phase at a low temperature. As we
increase the temperature, a ``fragile'' superfluid phase, where pairing is
assisted by a temperature induced smearing of the Fermi surfaces, reappears in a
window of temperatures.

\begin{center}
\begin{figure}[h!]
\includegraphics[width=4.5in,angle=0]{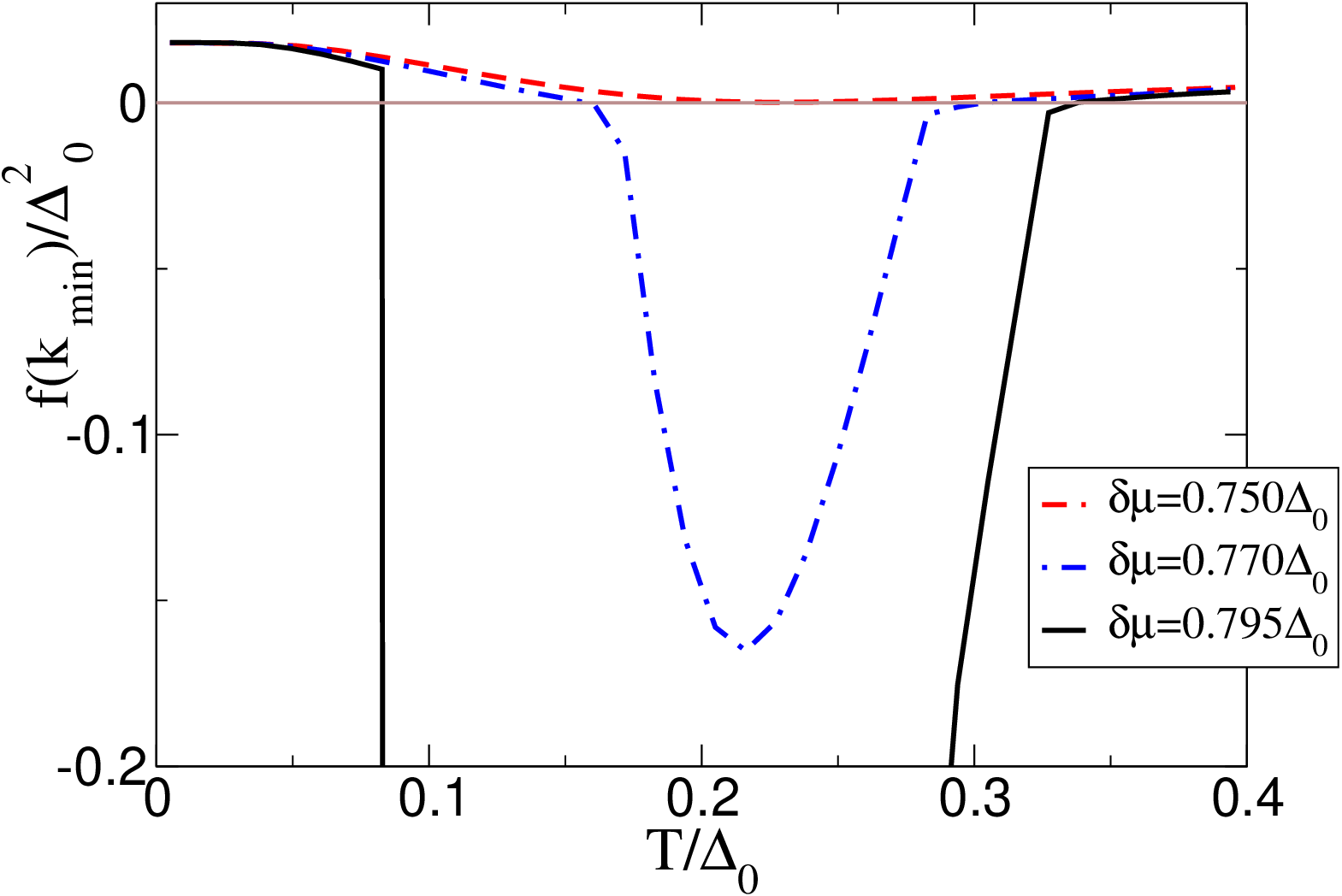}
\caption{$f_{\rm{min}}$ as a function of $T$ at three values of $\dmu$, 
$\dmu=0.750\Delta_0$, $\dmu=0.770\Delta_0$ and $\dmu=0.795\Delta_0$.
We take $g=-0.72$ as before. Looking first at $\dmu=0.795\Delta_0$, 
$f_{\rm{min}}$ is clearly negative in the unpaired region,
corresponding to $T/\Delta_0\in(0.08,0.22)$. It is also negative in the ``fragile''
superfluid region lying in $T/\Delta_0\in(0.22,0.33)$. Now considering
$\dmu=0.770\Delta_0$, we find that the homogeneous condensate is unstable 
in $T/\Delta_0\in(0.16,0.29)$. Finally for $\dmu=0.750\Delta_0$ (and lower), we find that
the homogeneous condensate is stable for all $T$. These three curves form the
basis of the thick dashed lines (blue online) in Fig.~(\ref{loff-pt}).} 
\label{instability}
\end{figure}
\end{center}

 As we shall see below, analysis of fluctuations about the mean field will tell us
that this ``fragile'' superfluid may give way to an inhomogeneous
phase~(Figs.~(\ref{loff-pt},~\ref{instability})).  For $\dm/\Delta_0=0.750$,
$0.770$ and $0.795$, the temperatures for which the homogeneous condensate is
unstable is shown by the dotted lines (blue online) in Fig.~(\ref{loff-pt}).

  If this mean field picture is taken as it is (and ignoring the inhomogeneous
phases for a moment), this structure of the phase diagram could be observed in cold
atomic traps. As we go away from the center of the trap, the effective chemical
potential and therefore $\Delta_0$ decreases.  Since $T$ and $\dmu$ are constant
across the trap, this implies that $T/\Delta_0$ and $\delta\mu/\Delta_0$
increases. The ``fragile'' superfluid region can be probed by tuning the
parameters, namely the number of the two species in the trap $N_1$ and $N_2$,
the scattering length $a$, and the temperature $T$, so that for a given trap
geometry, we pass through the ``fragile'' superfluid region as we go from the
center of the trap to the outside.

\begin{center}
\begin{figure}[h!]
\includegraphics[width=4.5in,angle=0]{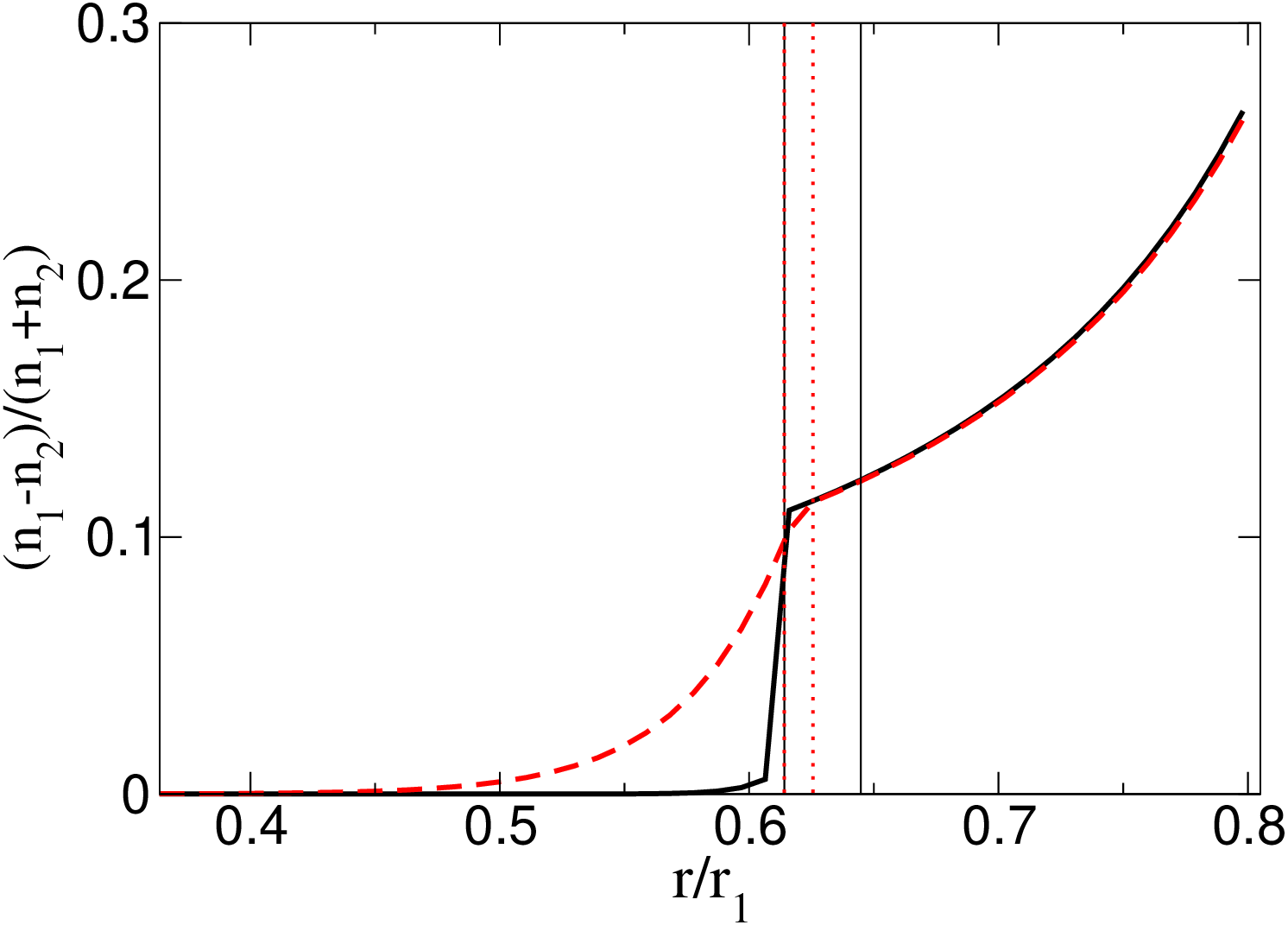}
\caption{(colored online) Polarization as a function of the $r$ plotted in units
of $r_1$. The bold line (black online) corresponds to $T=0.1\delta\mu$ and the
dashed line (red online) to $T=0.36\dmu$. Also shown are regions where the
system is unstable to the growth of inhomogeneities. The thin continuous lines
(black online) specify this region for $T=0.1\delta\mu$ and the thin dashed
lines (red online) mark this region for $T=0.36\delta\mu$. The lower boundary
marked by a thin dotted line, coincides with the left boundary of the thin
continuous line.  The upper boundary, however, appears at a larger $r$ for the
continuous line, telling us that the window of inhomogeneous pairing is wider
at lower temperature. For $T=0.1\dmu$ the total value of $N$ in the trap is
$3.21\times10^{(+7)}$ and the value of $\delta N$ is $1.26\times10^{(+6)}$.  For
$T=0.36\dmu$ the total value of $N=3.20\times10^{(+7)} $ and the value of
$\delta N = 1.53\times10^{(+6)}$. } 
\label{trap}
\end{figure}
\end{center}

 In Fig.~(\ref{trap}), we plot profiles of the polarization $p=(n_1-n_2)/(n_1+n_2)$ 
as a function of $r$, where $r$ is the distance from the center of
the trap and $n_1$ and $n_2$ are the number densities of the two species. The
trap is taken to be spherically symmetric, with a harmonic trapping angular
frequency $\omega=100\;{\rm{rads/s}}$~\cite{Partridge:2006a,Shin:2006,
Partridge:2006b,Shin:RF2007,shin:2007}. The harmonic potential can be written as
$V(r)=1/2\times\hbar\omega \times(r/r_0)^2$ with $r_0=\surd(\hbar/(m\omega))$,
where $m$ is the mass of $^6$Li atoms. In natural units, $\omega = 2.599\times
10^{(-10)}\;{\rm{(eV)}}$, $m=5.61\times10^{(+9)} \;({\rm{eV}})$ and
$r_0=0.82816\;({\rm{eV}})^{-1}$.

 With the geometry of the trap specified, we now try to tune the parameters of
the experiment $N=(N_1+N_2)/2$, $\delta N=(N_1-N_2)/2$,
$a$ and $T$ and try to identify the region where there is a
phase transition from the superfluid phase to the unpaired phase. This is interesting 
because of two reasons. Firstly, since the polarization 
rises rapidly at this phase transition boundary, this boundary can be recognized
in the trap by observing the polarization as a function of $r$. Secondly, the
inhomogeneous phases are likely to be found near this boundary at low
temperatures, and we would like to estimate the range of $r$ where these phases
are likely to exist. We choose $a=-0.0205\;({\rm{eV}})^{-1}$, for which the
atomic system, especially at the center of the trap, can not be considered weakly
coupled, and hence our approximations may not be quantitatively accurate near
the center of the trap. But the large coupling amplifies the features in the
polarization profile that we are looking for.  We use these features to deduce
qualitative conclusions which may be more general than our approximations.  

 Although the true experimental variables are $N$ and $\delta
N$, we find it more convenient to fix the values of the average chemical
potential at the center of the trap, $\mu_0$, and the chemical potential
splitting $\delta\mu$, and calculate $N$ and $\delta N$ in terms of these. The
average chemical potential at a point $r$ in the trap is $\mu(r)=\mu_0-V(r)$.
Plotted in Fig.~(\ref{trap}) is $p$ as a function of $r$ for
$\mu_0=1.314\times10^{(-7)}\;{\rm{(eV)}}$ and 
$\dmu=1.2\times10^{(-8)}\;{\rm{(eV)}}$ for two different temperatures,
$T=0.1\dmu$ and $T=0.36\dmu$. We focus on the region near the place where we
observe the jump in the value of the polarization. The radius $r$ has been
plotted in units of $r_1$ defined as the radius where the effective chemical
potential of $\psi_1$ species is zero, i.e.
$r_1=r_0\surd(2(\mu_0+\dmu)/\omega)=33.22r_0$.

 We see that, as expected, the jump in the polarization at the phase transition boundary
becomes less sharp as we increase the temperature. We also mark the region where
the system is unstable to the formation of inhomogeneous phases. We also see that
the range of $r$ where inhomogeneity may develop, decreases in size as we
increase the temperature.

  But we expect that this mean field picture will be modified due to quantum corrections. One
thing to note from the double-dot-dashed curve (red online)
in~Fig.~(\ref{OmegaDeltaplot}) is that the free energy curve as a function of
$\Delta$ is very shallow. The mass of the fluctuations in the magnitude of the
condensate is therefore very small and therefore we expect quantum corrections
to be important in this ``fragile'' superfluid state. 

 The role of quantum corrections can be understood by considering the effective
action describing the fluctuations of the two fermion condensate about the mean
field value. If the system is stable to small fluctuations of the condensate, 
we may calculate their contribution to the free energy and see if they increase
the mean field value of the free energy or decrease it. Recently,
the effect of fluctuations on the free energy function has been calculated for
$T=0$ and $\dmu=0$~\cite{diener-2007}. These calculations suggest that the mean
field calculations overestimate the value of $\Delta_0$ as well as the value of
$n$ in the superfluid state.

 The first step in such a calculation is to find the effective potential describing
the dynamics of the fluctuation field and seeing if the system is stable
to small fluctuations in the mean field value. In the following, we analyze the
stability of the ``fragile'' superfluid to fluctuations that depend on position
and are purely imaginary for real $\Delta$. I.e., we consider a condensate of form 
\begin{equation}
\langle\psi_\alpha({\bf{r}})\psi_\beta({\bf{r}})\rangle =
\frac{1}{\lambda}\epsilon_{\alpha\beta}\bigl(\Delta + i\eta(\bfr)\bigr)\label{imaginary fluctuations}\;,
\end{equation}
where $\Delta$ and $\eta$ are both real.  The reason for considering purely
imaginary fluctuations (eq.~(\ref{imaginary fluctuations})) is that it is clear
from the free energy curve at $T=0.24\Delta_0$ in~Fig.~(\ref{OmegaDeltaplot}) that
the free energy is at a local minimum at the mean field value $\Delta$, meaning
that the system is stable to spatially uniform variation in the magnitude of the
condensate. One could potentially consider position dependent fluctuations in
the magnitude of the condensate but we only consider real $\eta(\bfr)$ 
in eq.~(\ref{imaginary fluctuations}), which is related to position 
dependent fluctuations in the phase of the
condensate. The effective potential describing the dynamics of $\eta$ can be
found in a Ginzburg-Landau expansion in the same way as was done in
section~\ref{sec:inhomogeneous}, where
we considered inhomogeneous fluctuations about the normal phase. The only
difference now is that the background has a non-zero condensate value, $\Delta$.
The final answer has the form,
\begin{equation}
\int d^3\bfr ( \Omega(\Delta(\bfr))-\Omega_N ) = \threeier{\bfk}
\eta(\bfk)\eta(-\bfk)\Bigl\{f_{\Delta}(|\bfk|) \Bigr\}~\label{GL about Delta}
\end{equation}
with,
\begin{equation}
f_\Delta(|\bfk|) =\frac{1}{4}\threeier{\bfp} 
\Bigl\{\frac{\bigl(\tilde{\xi}(\bfp+\bfk)-\tilde{\xi}(\bfp)\bigr)^2}
{\bigl(\tilde{\epsilon}(\bfp+\bfk)^2-\tilde{\epsilon}(\bfp)^2\bigr)}
\Bigl[\frac{g(\tilde{\epsilon}(\bfp))}{\tilde{\epsilon}(\bfp)}
-\frac{g(\tilde{\epsilon}(\bfp+\bfk))}{\tilde{\epsilon}(\bfp+\bfk)}\Bigr]\Bigr\}\;.
\end{equation}
The form in~eq.~(\ref{GL about Delta}) closely resembles the form in~eq.~(\ref{GL
about zero}), except now $\alpha=0$, which can be easily understood as the
consequence of the fact that a position independent fluctuation in the phase of
the condensate can not change the free energy.

 In~Fig.~(\ref{instability}) we plot the minimum value of $f_{\Delta}(k)$ (minimized
over $k=|\bfk|$) as a function of $T$ at $\dmu/\Delta_0=0.795$, $0.770$ and
$0.750$ for $g=-0.72$.  For $\dmu/\Delta_0=0.795$, we see that the minimum
value, $f(k_{\rm{min}})$, is negative in the ``fragile'' superfluid region lying
in $T/\Delta_0\in(0.22,0.33)$, indicating that it is unstable with respect to
developing a position dependent phase modulation. We
see~(Fig.~(\ref{instability})) that as we decrease $\dmu$, the homogeneous
superfluid region becomes stable for all $T$, as it should.

 In light of this instability of the ``fragile'' superfluid, our results of the
polarization profiles for the atomic trap may need to be modified. To find the
correct profile, we will need to calculate the number densities of the two
species in an inhomogeneous condensate more accurately than we have done above.
We leave this calculation for future work. Another issue that warrants further
investigation is the role of polarization effects (also known as the
Gorkov-Barkhudarov corrections) that can reduce the gap. In weak coupling,
medium polarization results in screening the interactions and reduces the gap
and the critical temperature by a factor of $(4e)^{(1/3)}\sim2.2$ when
$\delta\mu=0$ \cite{Gorkov:1961,Heiselberg:2000}. If we were to incorporate this
reduction by modifying the four-fermion coupling in the pairing channel to mimic
the Gorkov suppression at $T=0$ and $\dmu=0$, it is easy to argue that the phase
boundary (shown in Fig.~\ref{homogeneous-normal}) would simply be scaled by the
same reduction factor $2.2$ both in temperature and $\delta\mu$. However, a
systematic calculation of these medium effects at moderate coupling is
challenging and preliminary investigations suggest that the gap is not as
strongly suppressed ~\cite{PhysRevC.63.044310}.  Further, even in weak coupling
the nature of this suppression is not understood at finite population imbalance.
We leave a systematic calculation of how Gorkov corrections evolve with coupling
strength and  $\dmu$ for future work.

Despite these aforementioned caveats, we reiterate that our investigation here
points to several interesting qualitative features that arise only when
Hartree corrections are included. In particular we see that the region in phase
diagram susceptible to gradient instabilities is moderately enhanced by these
corrections. The shape of the normal-superfluid phase boundary is shown to
depend on the nature of these mean field energy shifts since they differ in the
two phases in the vicinity of the first-order transition. In particular the
Clogston-Chandrashekar point, which is given by $\delta \mu = \Delta/\sqrt{2}$,
is shifted to higher values of $\delta \mu$ by these corrections.


\section{Acknowledgments}
RS thanks Krishna Rajagopal for discussion and suggestions. The authors also
acknowledge discussions with Michael Forbes, Joe Carlson, Mark Alford and Alex Gezerlis. This
research was supported by the Dept. of Energy under contract W-7405-ENG-36 and
by the LANL/LDRD Program.

\bibliography{phases}

\end{document}